\documentclass[aps,prl,twocolumn,groupedaddress]{revtex4}

\usepackage{graphicx}
\usepackage{bm}
\usepackage{color}
\usepackage{amsmath}
\usepackage{soul}

\begin{document}

\def\mos2{$\mathrm{Mo S_2}$}
\def\vecd{\mathbf{\vec d}}

\title{Spin- and valley-polarized transport across line defects in monolayer $\mbox{\boldmath MoS$_2$}$}

\author{Artem Pulkin}
\affiliation{Institute of Theoretical Physics, Ecole Polytechnique F\'ed\'erale de Lausanne (EPFL), CH-1015 Lausanne, Switzerland}

\author{Oleg V. Yazyev}
\affiliation{Institute of Theoretical Physics, Ecole Polytechnique F\'ed\'erale de Lausanne (EPFL), CH-1015 Lausanne, Switzerland}

\date{\today}

\def\artem#1{\textcolor{red}{#1}}
\def\oleg#1{\textcolor{blue}{#1}}

\begin{abstract}
We address the ballistic transmission of charge carriers across ordered line defects in monolayer transition metal dichalcogenides. Our study
reveals the presence of a transport gap driven by spin-orbit 
interactions, spin and valley filtering, both stemming from a simple picture of spin and momentum conservation, as well as the 
electron-hole asymmetry of charge-carrier transmission. 
Electronic transport properties of experimentally observed ordered 
line defects in monolayer \mos2, in particular, the vacancy lines and
inversion domain boundaries, are further
investigated using first-principles Green's function methodology. Our calculations demonstrate the possibility of achieving nearly complete spin polarization of charge carriers in 
nanoelectronic devices based on engineered periodic line defects
in monolayer transition metal dichalcogenides, thus suggesting a 
practical scheme for all-electric control of spin transport.
\end{abstract}

\pacs{
73.63.-b,	
72.25.-b,	
75.70.Tj,	
61.72.Mm	
}

\maketitle

\noindent

The spin and valley degrees of freedom 
of charge carriers are being actively considered as a means of
extending the capabilities of present-day charge-based electronics.
A major challenge on the way to designing practical devices for generating, manipulating and 
detecting spin- and valley-polarized charge carriers lies in finding novel physical phenomena and suitable materials that will be utilized in such devices. 
Two-dimensional (2D) materials such as graphene, and more recently the transition metal dichalcogenides, appear to offer a number of valuable properties
for the emerging fields of spintronics \cite{wolf_spintronics:_2001,zutic_spintronics:_2004,chappert_emergence_2007,fert_nobel_2008}  and valleytronics \cite{rycerz_valley_2007,Xiao07}.  

\begin{figure}[b]
\includegraphics[width=8.6cm]{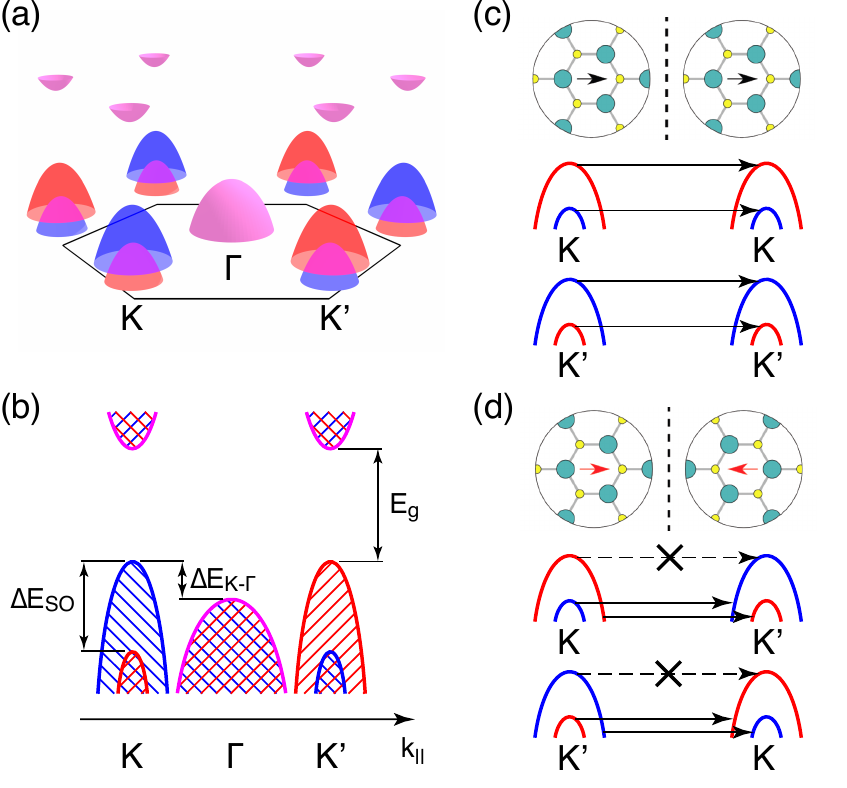}
\caption{
(a) Schematic illustration of the electronic band structure of
monolayer TMD materials showing the spin-split valence bands at points
$K$ and $K'$ of the Brillouin zone and nearly spin-degenerate valence band at $\Gamma$ and conduction bands at points $K$ and $K'$. The spin-up and spin-down bands are color coded (red and blue).
(b) Projection of the 2D band structure shown in (a) onto the direction parallel to one of the real-space lattice vectors.
(c),(d) Allowed transmission channels for the hole charge carriers in
the $K$ and $K'$ valleys across line defects that preserve lattice
orientation and inversion domain boundaries, respectively.
The insets show the orientation of the crystalline lattice in the two 
domains separated by the line defect (dashed line). 
}
\label{fig:1}
\end{figure}

The family of layered transition metal dichalcogenides (TMDs) $MX_2$ 
($M$ = Mo, W; $X$ = S, Se) has attracted considerable attention 
as prospective materials for next-generation electronics \cite{radisavljevic_single-layer_2011,radisavljevic_integrated_2011,Wang12} and photovoltaics \cite{bernardi_extraordinary_2013}.
Single layers of the $2H$-phase TMDs are direct band gap semiconductors 
with valence and conduction bands located at the inequivalent $K$ and $K'$ points of the Brillouin zone [see Fig.~\ref{fig:1}(a) for a schematic illustration of the band structure]
\cite{splendiani_emerging_2010,mak_atomically_2010,jin_direct_2013,zhang_direct_2014}.
Single-layer TMDs lack inversion symmetry, hence a spin-orbit interaction lifts the spin degeneracy of the bands. The effect of a spin-orbit interaction is particularly pronounced in the valence band, giving rise to spin splittings of 0.15--0.46~eV across the family of single-layer TMDs \cite{zhu_giant_2011}. A combination of the two
properties mentioned above results in an intrinsic spin-valley coupling 
of the hole charge carriers \cite{xiao_coupled_2012}, thus making these materials an appealing choice for beyond-electronics applications. However, the potential of this novel physical phenomenon for prospective technological applications is far from being fully explored. While valley polarization of charge carriers by optical means has been recently reported \cite{zeng_valley_2012, mak_control_2012, cao_valley-selective_2012}, neither practical all-electric schemes for generating polarized charge carriers nor spin transport phenomena in monolayer TMDs have been demonstrated so far.

In this Rapid Communication, we propose a simple approach for generating spin- and valley-polarized charge carriers by means of transmission across periodic line defects in monolayer TMDs. Our study also reveals
a number of transport phenomena in these materials, such as the suppression of hole transmission across inversion domain boundaries driven by spin-orbit splitting of the valence band.
Our predictions, which are based on a simple and intuitive picture of spin and momentum conservation, are further supported by the results of first-principles 
Green's function calculations, which reveal the quantitative aspects of spin- and valley-polarized transport across
representative line defects in monolayer \mos2 that have been observed experimentally and can be engineered in a controlled manner. 


We will first discuss the general phenomenology without making recourse to 
any particular member of the family of TMD materials or to a specific 
defect structure. It is, however, worth stressing that line defects in
TMDs, such as grain boundaries and inversion domain boundaries (equivalent to $60^\circ$ grain boundaries), tend to exhibit well-ordered periodic structures \cite{zhou_intrinsic_2013, van_der_zande_grains_2013,  najmaei_vapour_2013, Yazyev14}. The periodicity vector $\mathbf{d}$ of a one-dimensional defect 
is defined by the commensurability condition, that is, by matching 
two translational vectors in the ``left'' and ``right'' domains separated by the defect \cite{yazyev_electronic_2010}
\begin{equation}
\mathbf{d} = n_{\mathrm L} \mathbf{a}_{1,{\rm L}} + m_{\mathrm L} \mathbf{a}_{2,{\rm L}} = n_{\mathrm R} \mathbf{a}_{1,{\rm R}} + m_{\mathrm R} \mathbf{a}_{2,{\rm R}},
\label{commensurability}
\end{equation}
where ($\mathbf{a}_{1,{\rm L}}$, $\mathbf{a}_{2,{\rm L}}$) and ($\mathbf{a}_{1,{\rm R}}$, $\mathbf{a}_{2,{\rm R}}$) are the lattice vectors of the ``left'' and ``right'' domains, respectively, and 
$(n_{\mathrm L}, m_{\mathrm L})$, $(n_{\mathrm R}, m_{\mathrm R})$ are pairs of integers.


Figure~\ref{fig:1}(b) shows the band structure of a monolayer TMD 
projected onto the direction of momentum parallel to the defect, 
$k_{||}$, when this defect is oriented along one of the lattice 
vectors [or, more generally, when $(n_{\mathrm L(\mathrm R)}-m_{\mathrm L(\mathrm R)}) \mod 3 \neq 0$]. In this situation, which is most often observed in experiments, the two valleys at points $K$ ($\nu = +1$) and $K'$ ($\nu = -1$) of the Brillouin zone are separated in $k_{||}$. 
Once $k_{||}$ is conserved upon transmission, two possibilities can be realized. For defects characterized by
\begin{equation}
(n_{\mathrm L}-m_{\mathrm L})\mod 3 = (n_{\mathrm R}-m_{\mathrm R})\mod 3 \neq 0,
\label{eq:rule0}
\end{equation}
such as the structures that do not change lattice orientation [e.g., $(n_{\mathrm L}, m_{\mathrm L})= (n_{\mathrm R}, m_{\mathrm R}) = (0, 1)$], valley indices are conserved upon transmission.
 Figure~\ref{fig:1}(c) shows the
allowed transport channels for hole charge carries 
when spin is additionally conserved upon transmission across such periodic 
defects. In this situation, spin and momentum conservation does not lead to transmission blocking at any charge-carrier concentration.

\begin{figure}
\includegraphics[width=8.6cm]{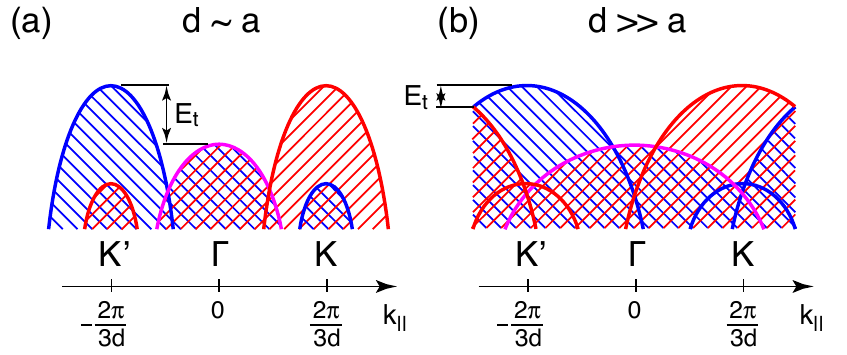}
\caption{
Illustration of the two scenarios governing the transport gap $E_{\rm t}$
across inversion domain boundaries for the hole charge carriers is
governed (a) by  $\Delta E_{K-\Gamma} < \Delta E_\mathrm{SO}$ or (b)
by the overlap between valleys $K$ and $K'$, which is realized for
sufficiently large line-defect periodicities $d$.
}
\label{fig:2}
\end{figure}

Figure~\ref{fig:1}(d) illustrates a different situation defined by 
\begin{equation}
0 \neq (n_{\mathrm L}-m_{\mathrm L})\mod 3 \neq (n_{\mathrm R}-m_{\mathrm R})\mod 3 \neq 0 ,
\label{eq:rule}
\end{equation}
in which the valley indices are exchanged upon crossing the defect.
This scenario is realized in inversion domain boundary defects, which can be defined 
by the lattice vectors
$\mathbf{a}_{1,{\rm L}} = -\mathbf{a}_{1,{\rm R}}$ and $\mathbf{a}_{2,{\rm L}} = -\mathbf{a}_{2,{\rm R}}$, and thus 
$(n_{\mathrm L}, m_{\mathrm L}) = (0, 1)$ and $(n_{\mathrm R}, m_{\mathrm R}) = (0, -1)$, in order to satisfy the commensurability condition (\ref{commensurability}). At low concentrations of hole charge carriers, this will result in a complete suppression of transmission 
if both spin and momentum are conserved. At larger charge-carrier concentrations, additional allowed channels will be involved once
the opposite-spin branches separated by spin splitting $\Delta E_\mathrm{SO}$ start being populated. In order to complete the picture of ballistic transmission across the inversion domain boundary defects, one needs to consider two more contributions: (i) transmission of the hole charge carriers populating the spin-degenerate valley at the $\Gamma$ point, which is separated from the valence band maximum by relatively small energies $\Delta E_{K-\Gamma}$ [see Fig.~\ref{fig:1}(b)], and (ii) the overlap of the $K$ and $K'$ valleys along $k_{||}$ at larger $d$, as illustrated in Fig.~\ref{fig:2}. Overall, one can expect a strong suppression of the 
transmission of hole charge carriers across the inversion domain boundaries, with the magnitude of the effective transport gap defined by
\begin{equation}
E_\mathrm{t} = \min \left [ \Delta E_\mathrm{SO}, \Delta E_\mathrm{K-\Gamma},  E_0 \frac{a}{d} \right ],
\label{eq:tg}
\end{equation} 
where $E_0 = h^2 /(72 m^\star a^2)$ is the characteristic energy of the valence band with an effective charge-carrier mass $m^\star$, and $a$ is the lattice constant.

\begin{figure*}[t]
\includegraphics[width=17cm]{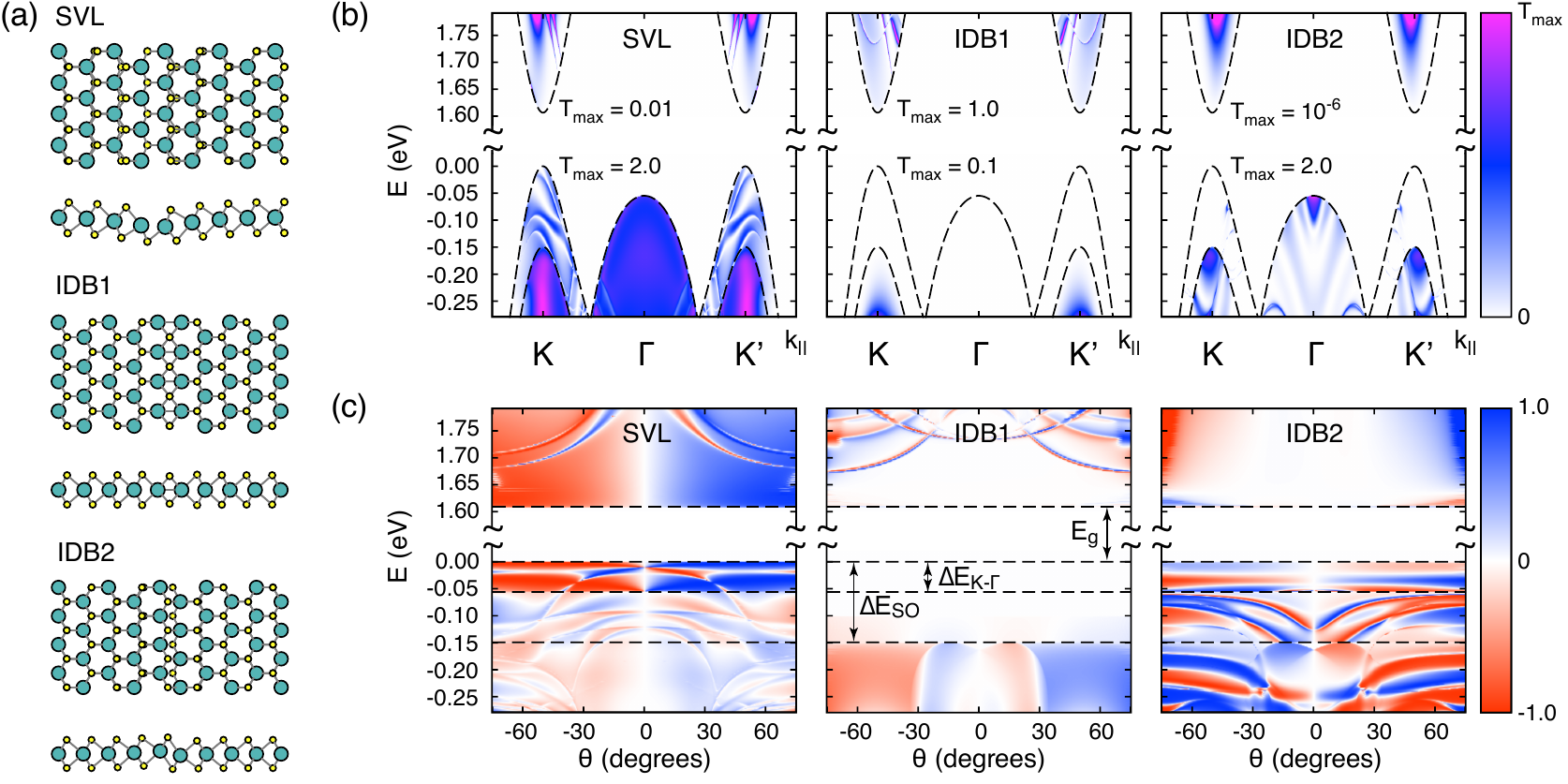}
\caption{
(a) Top and side views of the relaxed atomic structures of periodic line defects in monolayer \mos2, the single vacancy line (SVL), and two different structures of inversion domain boundaries (IDB1 and IDB2),
that have been observed experimentally \cite{komsa_point_2013,zhou_intrinsic_2013, lin_atomic_2014, Lehtinen15, Lin15}.
(b) Charge-carrier transmissions $T(E,k_{||})$ across the three investigated line defects in monolayer \mos2
as a function of energy $E$ relative to the valence band maximum and momentum $k_{||}$. 
The contours of the bulk bands projected onto the defect direction are shown as dashed lines. The transmissions of electron and
hole charge carriers are shown on different scales defined by the upper limit $T_{\rm max}$. 
(c) Spin polarization of the transmitted charge carriers $P_\sigma (E,\theta)$ as a function energy $E$ and incidence angle $\theta$ calculated for the three studied
line defects in monolayer \mos2.
Dashed lines reflect the positions of the band edges.}
\label{fig:3}
\end{figure*}

The fact that the two valleys are separated in $k_{||}$ in the 
considered scenarios will result in discrimination of the transmitted 
charge carriers with respect to their valley index $\nu$ and incidence 
angle $\theta \in \left ( -\pi/2, \pi/2 \right )$. This effect was 
originally predicted \cite{gunlycke_graphene_2011} for a particular 
experimentally observed line defect in graphene \cite{Lahiri10,Chen14}.  
We argue, however, that valley filtering is expected for 
any periodic line defect which leaves the two valleys separated in 
$k_{||}$ space since the following property holds for valley-resolved
transmissions,
\begin{equation}
T_\nu \left ( \theta \right ) = T_\nu \left ( k_{||} \right ) = T_{-\nu} \left ( -k_{||} \right ) = T_{-\nu} \left ( -\theta \right ) \neq T_{-\nu} \left ( \theta \right ),
\label{eq:sym}
\end{equation} 
for the incidence angle $\theta$ bijectively related to momentum $k_{||}$ in the one-dimensional (1D) Brillouin zone of the defect.
The corresponding valley polarization of the transmitted charge 
\begin{equation}
P_\nu \left (E, \theta \right ) = \frac{T_{\nu = +1} \left (E, \theta \right ) - T_{\nu = -1} \left (E, \theta \right )}{T_{\nu = +1} \left (E, \theta \right ) + T_{\nu = -1} \left (E, \theta \right )}
\end{equation}
is generally nonzero for any $\theta \neq 0$. At low concentrations, the valley polarization of hole charge carriers $P_\nu$ is equivalent to their spin polarization $P_\sigma$, due to the intrinsic spin-valley coupling in monolayer TMDs \cite{xiao_coupled_2012}.
The latter property, however, is different as spin is an intrinsic property of an electron, hence spin-polarized charge carriers can be injected into other materials. The considered defects 
can thus be used for controlling the spin polarization of charge 
carriers by all-electric means without the use of magnetic materials.

In order to investigate the quantitative aspects of the predicted spin- and valley-polarized transport phenomena, we perform first-principles quantum transport simulations using the Green's function technique with spin-orbit interactions
accounted for by using two-component spinor wave functions in combination with fully relativistic norm-conserving pseudopotentials (for a detailed description, see the Supplemental Material \cite{methodology}).
Without loss of generality, we further focus our attention on monolayer \mos2 as a representative member of the monolayer TMD materials family. In our 
density functional theory calculations, we obtain $E_{\rm g} = 1.61$~eV for the band gap. The calculated band offset between the $K$ and $\Gamma$ valence bands $\Delta E_\mathrm{K - \Gamma} = 0.055$~eV, while the spin splitting of the valence bands at $K$ points 
$\Delta E_\mathrm{SO} = 0.15$~eV, both in agreement with experiments and previous calculations \cite{jin_direct_2013,zhu_giant_2011}. The effective mass of hole charge carriers $m^\star = 0.57 m_e$ corresponds to the characteristic energy $E_0 = 0.73$~eV. As an example of a periodic 
line defect realizing the situation shown in Fig.~\ref{fig:1}(c), we 
consider the single vacancy line (SVL) defects reported by Komsa {\it et al.} \cite{komsa_point_2013}. Two different structures of 
inversion domain boundaries (IDB1 and IDB2), observed in Refs.~\onlinecite{zhou_intrinsic_2013, lin_atomic_2014, Lehtinen15, Lin15},
were considered as examples of systems realizing the second transport 
scenario shown in Fig.~\ref{fig:1}(d). 
All three defect structures are oriented along one of the lattice 
vectors (often referred to as the zigzag direction) and have the smallest possible
periodicity $d = a$.
Importantly, the above-mentioned ordered defects can be engineered with a fair degree of control at transmission electron microscopy conditions \cite{komsa_point_2013,lin_atomic_2014, Lehtinen15, Lin15}. The atomic structures of these line 
defects are shown in Fig.~\ref{fig:3}(a). 

Figure~\ref{fig:3}(b) shows the calculated charge-carrier transmissions 
$T(E,k_{||})$, across 
the investigated line defects in \mos2, as a function of energy $E$ relative to the valence band maximum and momentum $k_{||}$ parallel 
to the defect. One can immediately notice the predicted suppression of transmission across the inversion domain boundaries (IDB1 and IDB2) for the low-energy hole charge carriers. In monolayer \mos2, the transport gap is governed by the $K-\Gamma$ band offset rather than the spin splitting of the $K$ valley bands, that is, $E_{\rm t} = \Delta E_\mathrm{K - \Gamma}$ [cf.~Eq.~(\ref{eq:tg})]. In the case of the IDB1 defect, the transmission is strictly zero for $0 > E > -E_{\rm t}$, while for the 
IDB2 structure a residual transmission not exceeding 10$^{-3}$ was found
within the predicted transport gap. The latter is enabled by the 
spin-flip process due to out-of-plane bending at the defect line [Fig.~\ref{fig:3}(a)]. 
In general, $T(E,k_{||})$ shows strong
variations in both $E$ and $k_{||}$, with linelike suppression typical
of resonant backscattering on localized states hosted by the defects
\cite{Zou13,zhou_intrinsic_2013}.
A similar phenomenon was also reported for the charge-carrier transmission across a line defect in graphene \cite{Chen14}. 

Figure~\ref{fig:3}(c) shows the calculated spin polarization $P_\sigma (E, \theta)$ of transmitted charge carriers as a function of their 
energy $E$ and incidence angle $\theta$. The most prominent spin
polarization is found for the SVL defect at energies $0 > E > -\Delta E_\mathrm{K - \Gamma}$ where the hole charge carriers belong only to 
fully spin-polarized valleys $K$ and $K'$. For instance, at $E = -10$~meV, spin polarization achieves $P_\sigma = \pm 0.997$ at incidence angles $\theta = \pm 30^\circ$. 
Importantly, at these charge-carrier energies the transmissions are also high, of the order of 1. This combination of properties identifies the optimal conditions for using ordered vacancy
line defects in nanoscale spintronic devices. 
At $E < -\Delta E_\mathrm{K - \Gamma}$ the spin polarization dramatically reduces due to a large contribution to conductance of charge carriers in the spin-degenerate $\Gamma$ valley, which is also
characterized by large transmissions.
Interestingly, large values of $P_\sigma$ are  also found for the 
electron charge carriers, although spin-orbit effects in the conduction
band are generally weaker and have a complex character.
In this case, transmissions are found to be about two orders of magnitude lower. 

For the IDB1 defect we generally find a lower degree of $P_\sigma$ across
the relevant range of values of $E$ and $\theta$. The dominant contribution to the transmission of holes comes from conductance channels 
involving  $K$ and $K'$ valleys, 
as indicated in Fig.~\ref{fig:1}(d).
A sharp decrease of spin polarization within $\theta \approx \pm 30^\circ$ is due to the fact that only one of the two indicated channels is 
realized in this range of incidence angle values \cite{Koshino15}.
Specifically, transmission to the topmost valence band without a spin flip in this region is prohibited. 
The degree of spin polarization $P( E, \theta )$ of holes transmitted across IDB2 shows large variations with respect to energy $E$ and angle $\theta$, even in the energy range which corresponds to residual 
transmission enabled by the spin-flip process.
The sign of $P_\sigma$ at a constant incidence angle $\theta$ changes 
at $E \approx -20$~meV.
For $E < -\Delta E_\mathrm{K - \Gamma}$ multiple conductance channels compete, resulting in an irregular behavior of spin polarization.

\begin{figure}
\includegraphics[width=7cm]{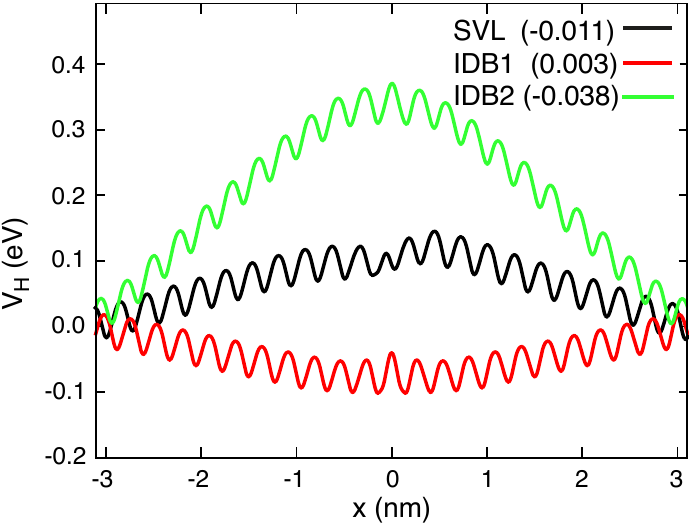}
\caption{
Self-consistent Hartree potential $V_{\rm H}$ averaged over the planes perpendicular to the transport direction for the three line-defect
models investigated in our work. The average $V_{\rm H}$ of the bulk monolayer \mos2 contacts is set to zero in all three cases.
The position of the defect is at $x = 0$. Calculated effective charges 
of defects per lattice constant length are given.}
\label{fig:4}
\end{figure}

It is interesting to note the pronounced electron-hole asymmetry of the ballistic conductances in all three considered cases 
[Fig.~\ref{fig:3}(b)]. The calculated transmissions of the hole charge 
carrier across the SVL and IDB2 line defects tend to approach their maximum 
nominal value of 2, which is the largest number of transmission channels within 
the investigated energy range. In contrast, the transmission of electron charge carriers is greatly suppressed across both defects. 
The opposite is true for the IDB1 line defect -- the transmission of low-energy electron charge carriers is generally higher, which was
also predicted for the same defect structure in MoSe$_2$ \cite{Lehtinen15}. We explain this behavior from the point of view of
electrostatic potential bending at the defect, which can be quantified
in terms of the self-consistent Hartree potential $V_{\rm H}$ within the scattering 
region obtained from our first-principles calculations (Fig.~\ref{fig:4}). For both the SVL and IDB2 line defects we observe 
upward potential bending which leads to a tunnelinglike transmission of 
electrons (hence their lower transmission), but not holes.
Moreover, the larger height of the potential barrier in the case of IDB2
is reflected in lower transmissions of low-energy 
electrons in comparison with the SVL defect.
The reversed behavior of charge carriers crossing the IDB1 defect 
is related to the downward potential bending.
The sign and magnitude of potential bending is defined 
by an effective charge localized on the defect (cf. Fig.~\ref{fig:4}), and a competing contribution due to the head-to-head change of polarization taking place upon crossing the IDB1 and IDB2 defects, as shown in the inset to Fig.~\ref{fig:1}(d). 
A similar polarization discontinuity has recently been predicted to occur at the interfaces created upon selective functionalization of 2D materials \cite{Gibertini14}.
The contribution of polarization discontinuity
dominates in the case of the IDB1 structure, thus leading to a downward potential bending, while the effective large negative charge in the
case of the IDB2 defect results in an upward potential bending.

In conclusion, our work reveals a number of transport phenomena
in the transmission of charge carriers across ordered line defects in monolayer \mos2 stemming from a simple and intuitive picture of spin and momentum conservation combined with strong spin-orbit effects in this two-dimensional material. The results are valid for other members of the monolayer TMD family of materials, but one can also
expect such phenomena to be observed in other semiconductors featuring strong spin-orbit interactions. Our work constitutes an important step towards understanding the transport properties of realistic samples of 
monolayer TMDs. Furthermore, we believe it also opens an avenue towards conceptually different nanoscale devices for electronics, and their extensions, such as 
spintronics and valleytronics. 
The energy of the charge carriers in such devices operated in the ballistic regime can 
be controlled by means of gating, while the angle dependence can be harnessed by positioning local contacts in predefined configurations relative to the line defect \cite{Yu11,Tsen12,Lee15}.
For the purpose of confirming the predicted filtering phenomenon,
coupled spin and valley polarizations of the charge carriers can be
detected using ferromagnetic contacts or by optical means \cite{zeng_valley_2012,mak_control_2012,Shan15}. For instance,
one can perform spatially resolved measurements of the circular polarization of light emitted upon electroluminescence \cite{Ross14} in devices containing line defects.
Practical devices relying on all-electric schemes could rather 
use components in the valve configuration, as suggested for a graphene-based valleytronic device \cite{Chen14}.
Considering the recently demonstrated long lifetimes of spin-polarized charge carriers in monolayer TMDs \cite{Yang15}, our work opens an avenue  
towards developing practical schemes for achieving all-electric control of spin transport in spintronic devices.


We thank G.~Aut\`es, F.~Gargiulo, and A.~Kis for discussions. This work was supported by the Swiss NSF (Grant No. PP00P2\_133552) and the ERC Starting grant ``TopoMat'' (Grant No. 306504). First-principles calculations have been performed at the Swiss National Supercomputing Centre (CSCS) under project s515.


\end{document}